\documentclass[
preprint,   
 amsmath,amssymb,
 aps,
]{revtex4-1}
\usepackage{graphicx}
\usepackage{dcolumn}
\usepackage{bm}
\usepackage{fix-cm} 
\usepackage{amsmath}
\usepackage{amssymb}
\usepackage{graphics}
\usepackage{pgfplots}
\usepackage{tikz}
\usetikzlibrary{decorations.pathmorphing}
\usetikzlibrary{decorations.markings}
\usetikzlibrary{decorations.pathmorphing}
\usetikzlibrary{decorations.pathreplacing}
\usetikzlibrary{patterns}
\usetikzlibrary{patterns.meta}
\usetikzlibrary{math}
\usepackage{caption}
\usepackage{ragged2e}
\usepgfmodule{oo}

\DeclareMathAlphabet{\mathsfbd}{T1}{\sfdefault}{\bfdefault}{\itdefault}
\SetMathAlphabet{\mathsfbd}{bold}{T1}{\sfdefault}{\bfdefault}{\itdefault}

\DeclareMathAlphabet{\mathsfit}{T1}{\sfdefault}{\mddefault}{\sldefault}
\SetMathAlphabet{\mathsfit}{bold}{T1}{\sfdefault}{\bfdefault}{\sldefault}

\DeclareMathAlphabet{\mathsfbdit}{T1}{\sfdefault}{\bfdefault}{\sldefault}
\SetMathAlphabet{\mathsfbdit}{bold}{T1}{\sfdefault}{\bfdefault}{\sldefault}

\begin{document}
\title{ Simultaneous Multiphoton-Multiatom Processes in Atomic Gases 
and Their Application in Enhancing Ultraweak Atomic Absorption Transitions}

\author{Yongle Yu }
\email{yongle.yu@wipm.ac.cn}

 \affiliation{ 
  Wuhan Institute of Physics  and Mathematics, Chinese Academy of Science,\\
 West No. 30 Xiao Hong Shan, Wuchang, Wuhan, 430071, China \vspace{2em}}

\begin{abstract}

We investigate simultaneous multiphoton-multiatom (MPMA) processes 
in atomic gases subjected to laser fields. Our study reveals that 
the  composite factor governing the transition rate of these processes can 
reach extraordinarily high magnitudes, with an intrinsic 
regulation mechanism causing the rate to exhibit near-saturation behavior. 
By integrating an MPMA process into an ultraweak atomic 
absorption transition, a substantial enhancement of the overall 
transition rate can be achieved. 
 This enhancement enables the detection of transitions that would 
 otherwise remain undetectable, thereby opening new avenues for 
 exploring ultraweak quantum phenomena in atomic systems.

\end{abstract}

\maketitle

Detecting ultraweak phenomena with extremely low 
probabilities is a fundamental challenge in 
contemporary physics, essential for  validating 
core theoretical postulates. Noteworthy among 
these endeavors are the detection of neutrinos \cite{neutrino} 
and the observation of atomic two-photon absorption 
events \cite{twophoton}. Neutrinos, distinguished by their exceptionally 
weak interactions with matter, require expansive detection volumes 
for observation. Similarly, atomic two-photon absorption—where 
an atom simultaneously absorbs two photons—occurs with 
negligible probability under natural conditions, but
becomes observable through laser-induced enhancements.
In this paper, we introduce a novel quantum amplification 
mechanism that has remained elusive until recently.
We examine a simultaneous multiphoton-multiatom (MPMA) 
process in an atomic gas \cite{WhiteAtomgasTlBa, NaAtoms, rios1980lineshape, andrews1983cooperative, 
 Nayfeh1984, Kim1998, Muthukrishnan2004, Zheng2013, twoMolecules, yu1}, demonstrating its capability 
to enhance ultra-weak atomic absorption transitions. Some
 fundamental mechanisms and unusual properties of the
 MPMA process have been recently elucidated in our work \cite{yu1}.

Consider a non-interacting atomic gas exposed to a laser field.
 This gas consists of two atomic species, labeled $A$ and $B$. (The introduction 
 of two distinct species here is primarily for formal convenience; a single-species 
 scheme can also be naturally implemented \cite{NaAtoms, yu1}.) For 
 simplicity, both $A$-species and $B$-species atoms are modeled as 
 two-level systems, where the upper level is accessible from 
 the lower level through an electric dipole transition. For 
 an $A$-species atom, let the lower level (ground state) energy 
 be $\varepsilon^a_g$ and the upper level (excited state) energy 
 be $\varepsilon^a_e$. The angular transition frequency is 
 given by $\omega_a = (\varepsilon^a_e - \varepsilon^a_g)/{\hbar}$, 
 where $\hbar$ is the Planck constant. Similarly, for a $B$-species atom, 
 the transition frequency is $\omega_b = (\varepsilon^b_e 
 - \varepsilon^b_g)/{\hbar}$, where $\varepsilon^b_e$ and $\varepsilon^b_g$ 
 represent the upper and lower level energies, respectively. 

Assume the angular laser frequency $\Omega_\mathfrak{L}$ is set to satisfy the
relation:
$m \hbar \Omega_\mathfrak{L} = \hbar  \omega_a + (m-1)\hbar \omega_b$,
where $m$ is an integer greater than one. 
In this scenario, when $\omega_a \neq\omega_b$, a single atom cannot 
be excited by absorbing
 a laser photon alone.
However, in a high-order Quantum Electrodynamics (QED) process, 
an  $m$-atom system comprising an $A$-species atom and 
$(m-1)$ $B$-species atoms, can be collectively excited when each 
atom absorbs a laser photon simultaneously. 
This phenomenon is known 
 as the multiphoton-multiatom excitation (MPMA) process. 
A schematic plot of a three-photon-three-atom process 
is depicted in Fig. \ref{fig_3p3a}. 

\tikzmath{ \bx=-1.2; \bz= 0.4; \bsz=0.5; \bsf=6;    
\wsf=\bsf+5; 
\wint= 0.12;  
\wx= -3.8; \wy=0.8; \wz= 0.7; 
\wsz=1.5;
\asz=0.07;
\rEx= 0.1; \Esz=1.1; \rEy= -1.4;  \rEz=0.2;\rint=2.2; 
\bEx= -0.1;  \bEy= 1.0; \bEz= -0.1;  \bint= 1.0; \brEyg=0.3; 
\Elabsz= 0.85; 
\atomrEgapx=1.2;\atombEgapx= 1.0; \atombExn=-0.18; 
\rbmix= 0.65; \rbmixn= 1-\rbmix;
\atomsize= 0.6;  
\pad= 0.05;
\seglen= 8pt; \amp=1.pt; 
\padint= 1.3pt;
 }

\tikzmath{ \arcR=40; \arcAng=5; \arcH=1.8;  \opacity=0.7; \ballR=0.1cm; 
   \centrX=4.4; \atomShift=0.9; \centrY=-0.9; \bEx=1;\rEx= -0.7;  \Elabsz= 0.85;
   \rEy= -0.03; \bEy=\rEy; \rint= 1.2;\bint=1.0; \bsz=0.5;  \pad=0.05; \seglen= 5pt; \amp=0.6pt;\Esz=0.7; 
\padint= 0.4pt; \wsz=0.6; \bEx= \rEx + \Esz+ 0.1; \bsf=0; \wint=0.02; \wx=0.24; \levelshift=-0.2; \wyinz=0.1;\ypanel=3.0;   \xpanel=5.0;
\photontraindx= 1.; }
\definecolor{LaserPink}{RGB}{241,184,186}
\definecolor{lightgreen}{RGB}{213, 245, 227} 
\definecolor{algeagreen}{RGB}{100, 233, 134}
 \pgfooclass{photonlineRatio}{

  \method apply(#1,#2,#3,#4,#5,#6,#7) {
   \draw[decorate,decoration={snake, segment length= #7*\seglen, amplitude=#7*\amp},color=#6](#1, #2)  -- +(#3,0);
  \draw[-stealth,color=#6](#1+#4,#2)  -- +(#5*1.5,0);
  
  }

 }
 
 \pgfooclass{photonlineRatioThick}{

  \method apply(#1,#2,#3,#4,#5,#6,#7) {
   \draw[thick, decorate,decoration={snake, segment length= #7*\seglen, amplitude=#7*\amp},color=#6](#1, #2)  -- +(#3,0);
  \draw[thick, -stealth,color=#6](#1+#4,#2)  -- +(#5*1.5,0);
  
  }

 }
 \pgfooclass{atomTransition}{
   \method two_levels(#1,#2,#3,#4,#5,#6,#7) {
   \draw[thick ,color=#6](#1, #2)  -- +(#3,0);
   \draw[thick ,color=#7](#1, #2+#4)  -- +(#3,0);
  }
   \method excitation(#1,#2,#3,#4,#5,#6){
    
   \draw[color=#6](#1+#3/2, #2)  circle(#5);
   \draw[dashed, color=#6, -stealth] (#1+#3/2, #2)--+(0,#4);
    }
     
   \method atom_frequency(#1,#2,#3,#4,#5,#6, #7){
       \node[color=#6, font=\tiny] at (#1+#3/2+4.5*\pad, #2+#4/2)  {#7};
       }
   \method photon_frequency(#1,#2,#3,#4,#5,#6,#7){
       \node[color=#6, font=\tiny] at (#1+#3/2+4.5*\pad, #2+#4/2)  {#7};
       } 
       
   \method photon_absorption(#1,#2,#3,#4,#5,#6) {
   \draw[dashed, color=#6] (#1, #2+#4)--+(#3,0);
   
   \draw[decorate,
 decoration={snake, segment length= \seglen, amplitude=\amp},color=#6](#1+#3/2, #2)  -- +(0,#4);
 
    }
}

\pgfooclass{atomgas}{
   \method apply(#1,#2,#3,#4,#5,#6,#7,#8) {
   \begin{scope}[]
   \draw [fill, rotate=#5, color= #8] (0,0) ellipse ({#1} and {#1*#2});
   \pgfmathsetseed{#4}
   \foreach \p in {1,...,700}
   {\tikzmath{\r= sqrt(rand+1.0)/1.414; \y=rand; \xx=#1*\r*sin(314*\y)*0.98; \yy=#1*#2*\r*cos(314*\y)*0.98;} 
     \fill[color=#7, rotate=#5](\xx,\yy ) circle (#3);
   }
    \fill[color=#6, rotate=#5](0,-0.2*#3*1.4) circle (#3*1.4);
   \end{scope}
   }
   
   \method corrlength(#1,#2,#3,#4,#5) {
   \begin{scope}[]
   \draw [fill, color=#4, opacity=#3 ] (0,0) circle ({#1});
   \draw[stealth-stealth](-#1,0)--(#1,0);
   \node[font=\bfseries,scale=0.8, color=#5] at (#1*0.4, #2) {$l_{mpma}$};
   \end{scope}
   }
   
   \method length_shrink(#1,#2,#3,#4,#5,#6) {
   \begin{scope}[]
   \draw [fill,  color= #5, opacity=#4 ] (0,0) circle ({#1});
    \draw[stealth-stealth, dashed, rotate=20, color=#6](-#1,0)--(#1,0); 
  
   \foreach \p in {1,...,6}
   {
   \draw[-stealth, rotate=60*\p, dash pattern={on 3*\padint off 3*\padint on 3*\padint }, color=#6](#1,0)--(#1*#2,0);
   
   }
   \node[font=\tiny, rotate=20, color=#6] at (35:#1*0.7) {$l^\Delta_{cor}$};
   \end{scope}
   }
   
}
\pgfoonew \myphotonlineRatioThick=new photonlineRatioThick()
\pgfoonew \myphotonlineRatio=new photonlineRatio()
\pgfoonew \myatomgas=new atomgas()  
\pgfoonew \myatomTransition=new atomTransition()
\pgfoonew \myatomTransition=new atomTransition()

\tikzmath{\atomtranx=0.0; \atomtrany= 0.0; \atomtranw=1.6; \atomtranh=1.8; \next= \atomtranw*1.2; \brratio=0.7;  \longline= 2.0; 
      \linelength=1.9; \large=1.5; \linexinc=0.10;  \lineyinc=0.15; \atAx= \wx+ 1.5* \longline*\wsz; \atgap= 0.7; \atAy= -16*\wint; \atBx= \atAx + 0.4*\atgap;
      \atBy= \atAy + 0.9*\atgap; \atCx= \atAx + 0.7*\atgap; \atCy= \atAy + 0.35*\atgap; \levelgap= 0.8; \atAlevelx= \atAx +1.0*\levelgap;
       \atAlevely= \atAy - 1.1*\levelgap; \atBlevelx= \atBx +0.5*\levelgap;
       \atBlevely= \atBy + 0.6*\levelgap; \atClevelx= \atCx + 1.5*\levelgap;
       \atClevely= \atCy + 0.9*\levelgap;\pad=0.05; \padint= 1.3pt; \rbfac=0.7; \lineshift= 3;}
\colorlet{unexcited-bcol}{green!20!blue!20}
\colorlet{unexcited-rcol}{gray!30!blue!10!}
\colorlet{gray2}{gray!70}
\colorlet{bcol}{green!20!blue!80}
\colorlet{rcol}{magenta!60!}
\definecolor{Laserblue_detune}{RGB}{20, 163, 199}
\definecolor{webblue}{RGB}{114,188, 212} 
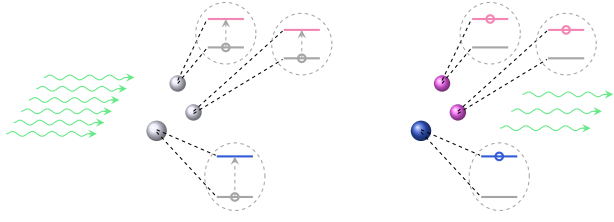
\begin{figure}
\captionsetup{width=.90\linewidth}
\begin{tikzpicture}
\begin{scope}[xshift=0, scale=1.0]

\myphotonlineRatio.apply(\wx - \linexinc, -3*\wint - \lineyinc, \linelength*\wsz, -2*\wint+\linelength*\wsz, \asz, algeagreen, \large)
\myphotonlineRatio.apply(\wx - 2*\linexinc, -3*\wint - 2*\lineyinc, \linelength*\wsz, -2*\wint+\linelength*\wsz, \asz, algeagreen, \large)
\myphotonlineRatio.apply(\wx, -3*\wint, \linelength*\wsz, -2*\wint+ \linelength*\wsz, \asz, algeagreen, \large)
\myphotonlineRatio.apply(\wx + \linexinc, -3*\wint + \lineyinc, \linelength*\wsz, -2*\wint+\linelength*\wsz, \asz, algeagreen, \large)
\myphotonlineRatio.apply(\wx + 2*\linexinc, -3*\wint + 2*\lineyinc, \linelength*\wsz, -2*\wint+\linelength*\wsz, \asz, algeagreen, \large)
\myphotonlineRatio.apply(\wx + 3*\linexinc, -3*\wint + 3*\lineyinc, \linelength*\wsz, -2*\wint+\linelength*\wsz, \asz, algeagreen, \large)
\
\node[circle, shading=ball, ball color=unexcited-rcol,  minimum size=2pt, scale=1.25*\atomsize,
text=white] (ball) at(\atAx,  \atAy){} ;
\node[circle, shading=ball, ball color=unexcited-rcol,  minimum size=2pt, scale=1.*\atomsize,
text=white] (ball) at(\atBx,  \atBy){} ;
\node[circle, shading=ball, ball color=unexcited-rcol,  minimum size=2pt, scale=1.*\atomsize,
text=white] (ball) at(\atCx,  \atCy){} ;

\myatomTransition.two_levels(\atAlevelx, \atAlevely, 0.3*\atomtranw, 0.3*\atomtranh, 0, gray2, bcol)
\draw[dash pattern={on \padint off \padint on \padint off \padint on \padint}](\atAx, \atAy-0.3*\pad)--(\atAlevelx-0.3*\pad ,\atAlevely +0.3*\pad );
\draw[dash pattern={on \padint off \padint on \padint off \padint on \padint}](\atAx, \atAy+ 0.3*\pad)--(\atAlevelx-0.3*\pad ,\atAlevely + 0.3*\atomtranh + 0.3*\pad );
\draw[dashed,dash pattern=on \padint off \padint, gray2, -stealth](\atAlevelx+ 0.5*0.3* \atomtranw, \atAlevely )--(\atAlevelx+ 0.5*0.3*\atomtranw, \atAlevely +  0.3*\atomtranh);
\draw[dashed,dash pattern=on \padint off \padint, gray2](\atAlevelx+ 0.5*0.3* \atomtranw, \atAlevely +  0.5*0.3*\atomtranh) ellipse ( 0.25*\atomtranw cm and 0.25*\atomtranh cm);
\draw[thick,gray2](\atAlevelx+ 0.5*0.3* \atomtranw, \atAlevely ) circle (0.05);

\myatomTransition.two_levels(\atBlevelx, \atBlevely, 0.3*\atomtranw, \rbfac * 0.3*\atomtranh, 0, gray2, rcol)
\draw[dash pattern={on \padint off \padint on \padint off \padint on \padint}](\atBx, \atBy-0.0*\pad)--(\atBlevelx-0.3*\pad ,\atBlevely - 0.3*\pad );
\draw[dash pattern={on \padint off \padint on \padint off \padint on \padint}](\atBx, \atBy+ 0.8*\pad)--(\atBlevelx-0.3*\pad ,\atBlevely + 0.3*\rbfac*\atomtranh - 0.3*\pad );
\draw[dashed,dash pattern=on \padint off \padint, gray2, -stealth](\atBlevelx+ 0.5*0.3* \atomtranw, \atBlevely )--(\atBlevelx+ 0.5*0.3*\atomtranw, \atBlevely +  0.3*\rbfac*\atomtranh);
\draw[dashed,dash pattern=on \padint off \padint, gray2](\atBlevelx+ 0.5*0.3* \atomtranw, \atBlevely +  \rbfac*0.5*0.3*\atomtranh) ellipse ( 0.25*\atomtranw cm and 0.25*\atomtranh *\rbfac*1.3 cm);
\draw[thick,gray2](\atBlevelx+ 0.5*0.3* \atomtranw, \atBlevely ) circle (0.05);

\myatomTransition.two_levels(\atClevelx, \atClevely, 0.3*\atomtranw, \rbfac * 0.3*\atomtranh, 0, gray2, rcol)
\draw[dash pattern={on \padint off \padint on \padint off \padint on \padint}](\atCx, \atCy-0.3*\pad)--(\atClevelx-0.3*\pad ,\atClevely - 0.3*\pad );
\draw[dash pattern={on \padint off \padint on \padint off \padint on \padint}](\atCx, \atCy+ 0.3*\pad)--(\atClevelx-0.3*\pad ,\atClevely + 0.3*\rbfac*\atomtranh - 0.3*\pad );
\draw[dashed,dash pattern=on \padint off \padint, gray2, -stealth](\atClevelx+ 0.5*0.3* \atomtranw, \atClevely )--(\atClevelx+ 0.5*0.3*\atomtranw, \atClevely +  0.3*\rbfac*\atomtranh);
\draw[dashed,dash pattern=on \padint off \padint, gray2](\atClevelx+ 0.5*0.3* \atomtranw, \atClevely +  \rbfac*0.5*0.3*\atomtranh) ellipse ( 0.25*\atomtranw cm and 0.25*\atomtranh *\rbfac*1.3 cm);
\draw[thick,gray2](\atClevelx+ 0.5*0.3* \atomtranw, \atClevely ) circle (0.05);

\end{scope}
\begin{scope}[xshift=100, scale=1.0]


\myphotonlineRatio.apply(\wx +\lineshift-1.5*\linexinc, -3*\wint- 1.5*\lineyinc , \linelength*\wsz, -2*\wint+ \linelength*\wsz, \asz, algeagreen, \large)
\myphotonlineRatio.apply(\wx +\lineshift, -3*\wint, \linelength*\wsz, -2*\wint+ \linelength*\wsz, \asz, algeagreen, \large)
\myphotonlineRatio.apply(\wx + 1.5*\linexinc +\lineshift, -3*\wint + 1.5*\lineyinc, \linelength*\wsz, -2*\wint+\linelength*\wsz, \asz, algeagreen, \large)
\
\node[circle, shading=ball, ball color=bcol,  minimum size=2pt, scale=1.25*\atomsize,
text=white] (ball) at(\atAx,  \atAy){} ;
\node[circle, shading=ball, ball color=rcol,  minimum size=2pt, scale=1.*\atomsize,
text=white] (ball) at(\atBx,  \atBy){} ;
\node[circle, shading=ball, ball color=rcol,  minimum size=2pt, scale=1.*\atomsize,
text=white] (ball) at(\atCx,  \atCy){} ;

\myatomTransition.two_levels(\atAlevelx, \atAlevely, 0.3*\atomtranw, 0.3*\atomtranh, 0, gray2, bcol)
\draw[dash pattern={on \padint off \padint on \padint off \padint on \padint}](\atAx, \atAy-0.3*\pad)--(\atAlevelx-0.3*\pad ,\atAlevely +0.3*\pad );
\draw[dash pattern={on \padint off \padint on \padint off \padint on \padint}](\atAx, \atAy+ 0.3*\pad)--(\atAlevelx-0.3*\pad ,\atAlevely + 0.3*\atomtranh + 0.3*\pad );
\draw[dashed,dash pattern=on \padint off \padint, gray2](\atAlevelx+ 0.5*0.3* \atomtranw, \atAlevely +  0.5*0.3*\atomtranh) ellipse ( 0.25*\atomtranw cm and 0.25*\atomtranh cm);
\draw[thick,bcol](\atAlevelx+ 0.5*0.3* \atomtranw, \atAlevely+ 0.3*\atomtranh ) circle (0.05);

\myatomTransition.two_levels(\atBlevelx, \atBlevely, 0.3*\atomtranw, \rbfac * 0.3*\atomtranh, 0, gray2, rcol)
\draw[dash pattern={on \padint off \padint on \padint off \padint on \padint}](\atBx, \atBy +0.0*\pad)--(\atBlevelx-0.3*\pad ,\atBlevely - 0.3*\pad );
\draw[dash pattern={on \padint off \padint on \padint off \padint on \padint}](\atBx, \atBy+ 0.8*\pad)--(\atBlevelx-0.3*\pad ,\atBlevely + 0.3*\rbfac*\atomtranh - 0.3*\pad );
\draw[dashed,dash pattern=on \padint off \padint, gray2](\atBlevelx+ 0.5*0.3* \atomtranw, \atBlevely +  \rbfac*0.5*0.3*\atomtranh) ellipse ( 0.25*\atomtranw cm and 0.25*\atomtranh *\rbfac*1.3 cm);
\draw[thick,rcol](\atBlevelx+ 0.5*0.3* \atomtranw, \atBlevely+\rbfac*0.3*\atomtranh ) circle (0.05);

\myatomTransition.two_levels(\atClevelx, \atClevely, 0.3*\atomtranw, \rbfac * 0.3*\atomtranh, 0, gray2, rcol)
\draw[dash pattern={on \padint off \padint on \padint off \padint on \padint}](\atCx, \atCy-0.3*\pad)--(\atClevelx-0.3*\pad ,\atClevely - 0.3*\pad );
\draw[dash pattern={on \padint off \padint on \padint off \padint on \padint}](\atCx, \atCy+ 0.3*\pad)--(\atClevelx-0.3*\pad ,\atClevely + 0.3*\rbfac*\atomtranh - 0.3*\pad );
\draw[dashed,dash pattern=on \padint off \padint, gray2](\atClevelx+ 0.5*0.3* \atomtranw, \atClevely +  \rbfac*0.5*0.3*\atomtranh) ellipse ( 0.25*\atomtranw cm and 0.25*\atomtranh *\rbfac*1.3 cm);
\draw[thick,rcol](\atClevelx+ 0.5*0.3* \atomtranw , \atClevely + 0.3*\rbfac*\atomtranh ) circle (0.05);

\end{scope}
\end{tikzpicture}
\caption{ \justifying Simultaneous joint excitation of a three-atom system, where each atom absorbs a laser photon.}
\label{fig_3p3a}
\end{figure}

In this $m$-atom system, the $(m-1)$ $B$-species atoms are 
distinctly labeled as $b_i \,(i = 1, 2, \dots, m-1)$. 
The Hamiltonian for the combined system, which includes these $m$ atoms 
and the laser field, is given by:
 \begin{equation}
{\hat H} =  \varepsilon^a_g |g_a\rangle \langle g_a| + \varepsilon^a_e |e_a\rangle \langle e_a|
   + \sum\limits_{1\leq i< m}( \varepsilon^b_g |g_{b_i}\rangle \langle g_{b_i}| + \varepsilon^b_e |e_{b_i}\rangle \langle e_{b_i}|)
   +  \hbar \Omega_\mathfrak{L} {\hat a^\dagger} {\hat a} + {\hat H_{int}},
\end{equation}
where $|g_a\rangle$ ($|e_a\rangle$) denotes the ground (excited) 
state of the $A$-species atom, and $|g_{b_i}\rangle$ ($|e_{b_i}\rangle$) denotes 
the ground (excited) state of the $b_i$ atom. Here, $\hat{a}$ ($\hat{a}^\dagger$) 
represents the annihilation (creation) operator of the laser photon, and $\hat{H}_{int}$ 
describes the coupling between the atoms and the laser field:
\begin{equation}   
{\hat H_{int}}= \hat{\mathbf{d}}_a \cdot {\mathbf {\hat E}} + \sum\limits_{i}\hat{\mathbf {d}}_{b_i} \cdot {\mathbf {\hat E}} ,
\end{equation}
where $\hat{\mathbf{d}}_a$ and $\hat{\mathbf{d}}_{b_i}$ are the 
dipole moments of the $A$-species atom and the $b_i$ atoms, respectively, 
and $\hat{\mathbf{E}}$ is the electric field operator of the laser.

Let the quantum state of the joint system with $m$ atoms in the 
ground states and $\scalebox{0.9}{$N_\gamma$}$ laser photons  be 
denoted as $|\Psi_i\rangle = |g_a\rangle |g_{b_1}\rangle |g_{b_2}
\rangle \dots |g_{b_{m-1}}\rangle |\scalebox{0.9}{$N_\gamma$}\rangle$ with an
energy $\varepsilon_{\Psi_{i}}= \varepsilon^a_g +  (m-1)\varepsilon^b_g + \scalebox{0.9}{$N_\gamma$} \hbar \Omega_\mathfrak{L}$. The quantum state
 with $m$ atoms being excited by absorbing $m$ photons 
 cooperatively is denoted as $|\Psi_f\rangle = |e_a\rangle |e_{b_1}\rangle 
 |e_{b_2}\rangle \dots |e_{b_{m-1}}\rangle |\scalebox{0.9}{$N_\gamma$}-m\rangle$ ($\scalebox{0.9}{$N_\gamma$} \gg m$). 
 The transition rate from $|\Psi_i\rangle$ to $|\Psi_f\rangle$ in the $m$-th order 
 perturbation is given by:
  \begin{equation}
 W_{mpma} = \frac {2\pi}{\hbar} |T_{fi}|^2 \rho(E_f)|_{E_f= \varepsilon_e^a + (m-1)\varepsilon_e^b }.
 \label{eq:w1}
 \end{equation}
 Here, $\rho(E_f)$ is the density of  states of the $m$-atom system 
 at energy $E_f$ \cite{densityStates}, and  
\begin{equation}
T_{fi}=\sum_{\kappa_1}\sum_{\kappa_2} \dots\sum_{\kappa_{m-1}}
\frac{\langle \Psi_f | {\hat H_{int}}| \Psi_{\kappa_{m-1}}\rangle \dots \langle \Psi_{\kappa_2}
{\hat H_{int}}| \Psi_{\kappa_{1}}\rangle \langle \Psi_{\kappa_1} | {\hat H_{int}}| \Psi_i\rangle} 
{ (\varepsilon_{\Psi_{\kappa_{m-1}}}  - \varepsilon_{\Psi_i})
\dots (\varepsilon_{\Psi_{\kappa_3}}  - \varepsilon_{\Psi_i})
( \varepsilon_{\Psi_{\kappa_2}} - \varepsilon_{\Psi_i}) (\varepsilon_{\Psi_{\kappa_1}} - \varepsilon_{\Psi_i})} ,
\end{equation}   
where $\Psi_{\kappa_j}$ is the virtual intermediate quantum state 
of the system at the $j$-th step in this $m$-th order 
process with an energy of $\varepsilon_{\Psi_{\kappa_j}}$, and the 
virtual transition path is $|\Psi_i\rangle \rightarrow |\Psi_{\kappa_1}\rangle
 \rightarrow |\Psi_{\kappa_2}\rangle \rightarrow \dots \rightarrow 
 |\Psi_{\kappa_{m-1}}\rangle \rightarrow |\Psi_f\rangle$. It 
 is clear  that $\Psi_{\kappa_j}$ corresponds to
 a quantum state with $j$ atoms  excited and with a laser
 photon number of $\scalebox{0.9}{$N_\gamma$}-j$; otherwise, its contribution to the transition vanishes.
 Notably, there are $m!$ possible excitation pathways corresponding to 
 different choices of intermediate states $|\Psi_{\kappa_1} \rangle ,|\Psi_{\kappa_2} \rangle,
 \dots ,|\Psi_{\kappa_{m-1}} \rangle $, leading to different sequences of transitions:
  $| \Psi_i\rangle \rightarrow |\Psi_{\kappa_1} \rangle \rightarrow
 |\Psi_{\kappa_2} \rangle \rightarrow \dots \rightarrow | \Psi_f\rangle$.
 
 A subtle aspect of this treatment is that
 $T_{fi}$ vanishes exactly at
 the resonant frequency $\Omega_\mathfrak{L}= \omega_a/m + (m-1) \omega_b/m $, due to
  quantum interference among different excitation pathways \cite{yu1}.
 However, this conclusion is fundamentally altered when some higher-order
  quantum electrodynamics (QED) effects are taken into account.
  These effects introduce finite widths for the excited levels.
  
To incorporate these effects, one adds an imaginary component to each of the excited-level energies:
   $ \varepsilon^a_e \rightarrow \varepsilon^a_e - i  \Gamma/2 $ and
   $ \varepsilon^b_e \rightarrow \varepsilon^b_e - i \Gamma/2 $, 
 where  $ \Gamma$ is the (natural) width of the excited levels,
assumed identical for both atoms for simplicity. 
With this modification,  $T_{fi}$ no longer vanishes, 
and quantum interference is found to contribute  a suppression factor to
  $|T_{fi}|^2$, expressed as  
  $ \frac{\Gamma^2/\hbar^2}{( \Omega_\mathfrak{L}- \omega_b)^2} [f_o(m)]^2$, where $f_o(m)$
 is a function of $m$ with values approximately in the range $(1/m, 1)$. It 
 is also noteworthy that  the exact cancellation of $T_{fi}$ in standard perturbation theory no longer holds
when $\Omega_\mathfrak{L}$ deviates from the exact resonant frequency. In such a detuned case, 
the MPMA process incorporates an inseparable subprocess of photon emission by the 
$m$-atom system. Again, quantum interference leads to a suppression 
factor analogous to that observed in the case of exact resonance \cite{yu1}.

Accounting for this quantum interference suppression factor, the
  transition rate
  can   be approximately estimated as:
 \begin{equation}
 W_{mpma} =  \frac{1}{{2}^{4m-1}\pi^{3m-1} \hbar} {(\Omega_\mathfrak{L}/\omega_a)}{(\Omega_\mathfrak{L}/\omega_b)^{m-1}} n_{\lambda^3}^m \frac{\Gamma^2 \gamma_a \gamma_b^{m-1} \Omega^m_\mathfrak{L}} {(\Omega_\mathfrak{L} - \omega_b)^{2m}} [f(m)]^{2m} \rho(E_f).
 \label{eq:w2} 
 \end{equation}
 In this expression, $\gamma_a$ represents a decay rate parameter of the $A$-species atom, defined as 
 $\gamma_a= {4 \alpha_e  \omega_a^3} |\langle e_a| {\mathbf{\hat{d}}_a}| g_a \rangle|^2 /{3 e^2 c^2}$, 
 where  $e$ is the charge of the electron, $c$ is the speed of light 
 and  $\alpha_e = {e^2}/{ 4\pi \hbar c \epsilon_0} \approx 1/137 $ is
  the fine-structure constant. Similarly, $\gamma_b= {4 \alpha_e  \omega_b^3} |\langle e_{b_0}| {\mathbf{\hat{d}}_{b_0}}| g_{b_0} \rangle|^2/{3 e^2 c^2}$ is a decay rate parameter of one $B$-species atom. The term  $n_{\lambda^3} $ represents the  number of laser photons in a volume of  
$\lambda^3 = (2\pi c/\Omega_\mathfrak{L})^3$, {\it i.e.}, $n_{\lambda^3} =  \scalebox{1.0}{$N_\gamma$}  \lambda^3 /V$ with $V$ being the
volume of the laser field. Finally, $f(m)$ is a function of $m$ with values roughly in the range $(1/m,1)$ \cite{f_absorb}.

In \cite{yu1}, we argued for the existence of a characteristic 
length, $l_{mpma}$, which plays a fundamental role in the cooperative excitation of an 
$m$-atom system. This excitation can only occur when the system's 
linear size is smaller than  $l_{mpma}$. A key 
 mechanism determining the 
value of $l_{mpma}$ arises  from the uncertainty principle in quantum mechanics.
In this $m$-th order process, while energy conservation 
is maintained between the initial and final quantum states of the 
joint system, all intermediate virtual quantum transitions
involve energy non-conservation.
The largest energy mismatch among the $m$ intermediate virtual quantum transitions, denoted $\Delta_\varepsilon$, 
is given by
  $  \hbar |\Omega_\mathfrak{L} - \Omega_a|$.
The uncertainty principle implies that these intermediate 
states can last a duration  $\Delta t = \hbar / 2\Delta_\varepsilon$. 
Consequently, a corresponding length, $l_{\Delta} \equiv c \Delta t$, emerges.
 It is natural to propose that 
 $l_{mpma} = \alpha l_{\Delta}$, where $\alpha$ is a constant of order 
  unity or less. 
 Some experimental evidence \cite{WhiteAtomgasTlBa, NaAtoms} 
 supporting this analysis of $l_{mpma}$
 is discussed in \cite{yu1}.
 
 To optimize the MPMA process for certain applications, it 
 could be crucial to achieve a large value of $l_{mpma}$ and a small 
 value of $\Delta_\varepsilon$. This can be more effectively 
 realized by utilizing two laser fields for the joint excitation 
 of the $m$-atom system \cite{yu1}. One laser, denoted  $\mathfrak{L_{1}}$, is 
 intended to excite the $A$-species atoms and has a frequency 
 $\Omega_\mathfrak{L_{1}}$ that differs from $\omega_a$ by a small
  amount $\delta_{a} = |\Omega_\mathfrak{L_{1}} - \omega_a|$. The 
  second laser, denoted $\mathfrak{L_{2}}$, has a frequency $\Omega_\mathfrak{L_{2}}$ close to $\omega_b$.
  The frequencies are configured to satisfy the relation:
  \begin{equation}
   \Omega_\mathfrak{L_{1}}+ (m-1) \Omega_\mathfrak{L_{2}}= \omega_a + (m-1) \omega_b.
   \end{equation}
   With this setup, the $m$-atom system can undergo joint 
   excitation by absorbing one photon from the $\mathfrak{L_{1}}$ laser 
   and $m-1$ photons from the $\mathfrak{L_{2}}$ laser. The transition rate 
   can be analyzed straightforwardly, taking a form similar to the right side 
   of Eq.~(\ref{eq:w2})  and involving the intensities of both lasers. In 
   this two-laser scenario, $\Delta_\varepsilon = \hbar \delta_a$, 
   which can be widely tuned.
  The potential smallness of $\delta_a$ is generally constrained by factors 
 such as the natural width of excited states and the thermal Doppler width of the
 atom gas. 
 In some cases, $\delta_a$ can reach the order of $2\pi \times 10$ GHz, allowing 
 $l_{mpma}$ to be as large as $3.0$ mm, which is around four orders of magnitude 
 larger than the wavelength of visible light. In the remainder of this paper,
 we will not differentiate whether the MPMA process involves one or 
 two laser fields, as the formalisms and estimations are quite similar.

According to Eq.~(\ref{eq:w2}), the transition rate for an $m$-atom 
system is inherently low, reflecting 
its nature as a high-order QED process. However, this rate applies
to a single m-atom system.  When considering the entire atomic gas, 
the overall transition rate for a specific atom can be significantly 
enhanced, driven by the key fact that a vast number of $m$-atom systems 
involving this atom engage in the MPMA process in parallel.

Consider an atomic gas with a linear size $l_{gas}$ and a homogeneous density 
of $B$-species atoms denoted by $\rho_b$.
The total number of $A$-species atoms is denoted by $\mathcal{N_A}$ and the total number 
of $B$-species atoms is $\mathcal{N_B}$ (assuming $\mathcal{N_A}< \mathcal{N_B}/m$). 
Select one $A$-species atom, labeled $A_o$, and denote the total number of $B$-species
atoms that can be jointly excited with the $A_o$ atom as
$N_{b\mathfrak{o}}$. If $l_{gas}<l_{mpma}$, then $N_{b\mathfrak{o}} = \mathcal{N_B}$
and the total number of $m$-atom systems involving the $A_o$ atom is 
$C^{N_{b\mathfrak{o}}}_{m-1} 
= \frac{N_{b\mathfrak{o}}(N_{b\mathfrak{o}}-1)\dots(N_{b\mathfrak{o}}-(m-1))}{(m-1)(m-2)\dots 1}
 \approx N_{b\mathfrak{o}}^{m-1}/(m-1)!$ . 
In the case where $l_{gas}>l_{mpma}$,  $N_{b\mathfrak{o}}$ can be approximated by 
$\rho_b l_{mpma}^3$, assuming a geometric factor of unity for simplicity.
The total number of  $m$-atom systems (with spatial extent less than  $l_{mpma}$)
 involving the $A_o$ atom in the
gas is roughly a fraction of $C^{N_{b\mathfrak{o}}}_{m-1} $, and we assume this fraction to 
be one for simplicity. (This fraction can be absorbed into an effective  $N_{b\mathfrak{o}}$,
 defined as the original  $N_{b\mathfrak{o}}$  multiplied by a factor  
 less than 1 but on the order of unity, which we approximate as 1.)
 According to quantum mechanics,  
all these $m$-atom systems engage in the process, and 
 the 
 overall transition rate for the $A_o$ atom is given by:
 \begin{equation}
 W_{a_o} \approx \frac{N_{b\mathfrak{o}}^{m-1}}{(m-1)!} W_{mpma}.
 \label{eq:w3}
 \end{equation}
This significant quantum enhancement  can be clearly illustrated  through 
the analysis of  relevant wavefunctions. 
For simplicity, consider a separate subsystem of the
 atomic gas composed of the  $A_o$ atom and $N_{b\mathfrak{o}}$ $B$-species atoms
with a linear size smaller than  $l_{mpma}$.  At $t=0$, with all atoms 
in their ground states, the wavefunction of the system 
(comprising  the $A_o$ atom, $N_{b\mathfrak{o}}$ $B$-species atoms, and the laser field) is 
 $\Psi_{ini}= |g_{a_o}\rangle |g_{b_1}\rangle |g_{b_2}\rangle \dots |g_{b_{N_{b\mathfrak{o}}}}\rangle | {\scalebox{0.9}{$N_\gamma$}} \rangle$. 
 For a specific choice of $m-1$ atoms out of 
 $N_{b\mathfrak{o}}$ $B$-species atoms 
indexed by $k_1, k_2, \dots, k_{m-1}$ ($k_1 < k_2 < \dots < k_{m-1}$),  the 
quantum state with the chosen  $m$-atoms being excited 
(by absorbing $m$ laser photons) is:
\begin{equation}
\Psi^{ex}_{k_1,k_2,\,\dots\,,k_{m-1}} = |e_{a_o}\rangle |e_{b_{k_1}}\rangle |e_{b_{k_2}}\rangle\dots |e_{b_{k_{m-1}}}\rangle |\Psi^b_{\overline{k_1,k_2,\,\dots \, ,k_{m-1}}}\rangle|{\scalebox{0.9}{$N_\gamma$} -m}\rangle
\end{equation}
Here, $|e_{b_{k_i}}\rangle$ (for $i=1,2,\dots,m-1)$ denotes the 
excited state of the $k_i$-th $B$-species atom, and $ |\Psi^b_{\overline{k_1,k_2,\,\dots\,,k_{m-1}}}\rangle $ 
represents the the quantum state of the remaining $B$-species atoms in their ground states.

At a small time $t$, the quantum state $\Psi(t)$ is given by:
\begin{equation}
\Psi(t) = c_0(t)\Psi_{ini} + \sum\limits_{k_1,k_2,\,\dots\,,k_{m-1}} c_{k_1,k_2,\dots,k_{m-1}}(t) \Psi^{ex}_{k_1,k_2,\,\dots\,,k_{m-1}} .
\end{equation}
According to perturbation theory, 
 $|c_{k_1,k_2,\dots,k_{m-1}}(t)|^2  \approx W_{mpma} t$. The probability 
 of the $a_o$ atom remaining unexcited is:
  \begin{equation}
P_{unex}(t) = 
|c_0(t)|^2 = 1 -  \sum\limits_{k_1,k_2,\,\dots \,,k_{m-1}}|c_{k_1,k_2,\,\dots \,,k_{m-1}}(t)|^2  \approx 1 - \frac{N_{b\mathfrak{o}}^{m-1}}{(m-1)!} W_{mpma} t .
\end{equation}
The rate of change of $P_{unex}(t)$ with respect to $t$ corresponds to
 the total transition rate.
At a time $\tau_{ex}$ 
corresponding to the inverse of  $W_{a_o}$,
$P_{unex}(\tau_{ex})$ deviates significantly from unity, indicating 
a high probability that the $A_o$ atom  has undergone an MPMA transition.

Combining Equations~(\ref{eq:w2}), and (\ref{eq:w3}), one finds,
\begin{equation}
W_{a_o} \approx \frac{1}{ 2^{4m-1}\pi^{3m-1} \hbar} n^m_{\lambda^3}\frac{\Gamma^2 \gamma_a \gamma_b^{m-1}\Omega^m_\mathfrak{L}} {(\Omega_\mathfrak{L}- \omega_b)^{2m}} [f(m)]^{2m} \frac{N_{b\mathfrak{o}}^{m-1}}{(m-1)!} \rho(E_f).
\label{eq:w_a_o}
\end{equation}
Consider a homogeneous gas system where 
 $N_{b\mathfrak{o}}\approx \rho_b l_{mpma}^3 $. 
According to Eq.~(\ref{eq:w_a_o}),
$W_{a_o}$ depends critically on $N_{b\mathfrak{o}}$, and thus on $l_{mpma}$.
In certain scenarios, $W_{a_o}$ can become exceptionally large when $l_{mpma}$ is significant.
 For instance, consider $l_{mpma} \approx l_{\Delta} \approx
 1.0$ mm, corresponding to $\Omega_{\mathfrak{L}} - \omega_a$ on the order of $2\pi \times 20$ GHz, 
 then $N_{b\mathfrak{o}} \approx 10^{12}$ 
with $\rho_b=10^{15}$ atoms/cm$^3$. 
Assuming $\gamma_a$ and $\gamma_b$ are on the order of $2\pi \times 10$ MHz, $\Gamma \approx \hbar \gamma_b$, 
$\rho(E_f) \approx 10^{-3}/\hbar \gamma_a$, and $n_{\lambda^3} 
\approx 10^{-2}$, $W_{a_o}$ exceeds $10^{20}$ s$^{-1}$ for $m=4$, an 
exceptionally high value. This suggests that a regulatory mechanism may 
be necessary to prevent $W_{a_o}$ from becoming excessively large.

In fact, there exists a fundamental mechanism which
 regulates the values of $l_{mpma}$
and $W_{a_o}$.
 The timescale $T_{a_o} \equiv 1/W_{a_o}$, represents the average 
duration for the  $A_o$  atom to complete the excitation transition. 
As this transition is part of a collective process involving 
an $m$-atom system, the entire system must transition within 
the same timescale $T_{a_o}$. As a result, the $m$-atom system 
should be confined within a distance smaller than
 $l_{a_o} \equiv c T_{a_o} = c/W_{a_o}$, ensuring that collective
 excitation complies with the principle that information 
 does not propagate faster than the speed of light. 
 Notably, most $m$-atom systems involving the $A_o$ atom 
 have a size close to the maximum allowed length $l_{mpma}$, while 
 those with significantly smaller sizes are negligible in number. This 
 introduces a further constraint: $l_{mpma} \leq \beta l_{a_o}$, where 
 $\beta$ is a constant less than unity. This condition is generally 
 satisfied when $W_{a_o}$ is small but becomes more restrictive as $W_{a_o}$ increases.

\begin{figure}
\begin{center}
\captionsetup{width=.86\linewidth}
\includegraphics{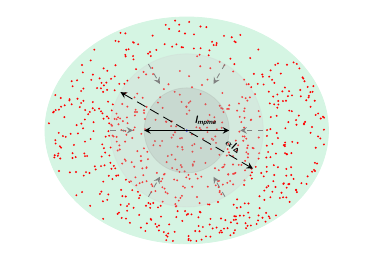}
\caption{\justifying In a dense gas, the characteristic length $l_{mpma}$ of 
an MPMA process can self-adjust to values below $\alpha l_\Delta$, ensuring
 that relativistic causality is not violated.}
 \label{fig_selftune}
 \end{center}
\end{figure}

This analysis leads to two scenarios for determining $l_{mpma}$:

 i) If $\alpha l_{\Delta} 
< \beta l_{a_o}$: $l_{mpma}$ can be directly set to $\alpha l_{\Delta}$,  and
 $W_{a_o}$ can be computed using Eq.~(\ref{eq:w_a_o}).
 
ii) If $\alpha l_{\Delta} > \beta l_{a_o}$: since $l_{mpma} \leq \beta l_{a_o} $ is required, 
$l_{mpma}$  can no longer equal  $\alpha l_{\Delta}$, Instead, it is adjusted 
to a reduced value (see Fig. \ref{fig_selftune}), which decreases $W_{a_o}$ 
and increases $l_{a_o}$, yielding the following self-consistent relationship:
\begin{equation}
  l_{mpma}= \beta l_{a_o} = \beta c /W_{a_o}.
\label{eq:reg}
  \end{equation} 
  Using Equations~(\ref{eq:w_a_o}) and (\ref{eq:reg}), along with
  $N_{b\mathfrak{o}} =  \rho_b l_{mpma}^3$, it becomes possible to consistently determine
$l_{mpma}$ and ${W_{a_o}}$.  This  regulating mechanism 
 significantly influences the value of
  $l_{mpma}$, ensuring a more physically 
   reasonable value  for 
 ${W_{a_o}}$ and preventing it  from becoming unnaturally large.
 
 \definecolor{lightgreen}{RGB}{213, 245, 227} 
\definecolor{lightgray}{RGB}{211, 211, 211}
\definecolor{gray2}{RGB}{137, 148, 153}   
\definecolor{atomcolorA}{RGB}{174, 214, 241} 
\definecolor{atomcolorB}{RGB}{162, 217, 206} 
\definecolor{algeagreen}{RGB}{100, 233, 134}
\definecolor{redgold}{RGB}{235, 84, 6}
\tikzmath{\latomgas=-0.3; \latomgaslcorr= 0.6; \ratomgas= 0.0; \atomgasxsz= 2; \atomgasratio= 0.8; \blaserx=-0.08; \blasery= -0.5;  \blaserl= 1.0; \blaserw= 0.2; \blaserfl= 1.0; \blaserarr= 1.2*\blaserw; \atomr=0.015;  \nodeyshift=0.13;
             \sensorx= 0.2; \sensory= 0.3; \sensorR=0.2;
             \shellin=1.5*0.4; \rain= 4;  \shellout= \shellin*1.15; \raout=3.7;}
\tikzmath{\atomtranx=0.0; \atomtrany= 0.0; \atomtranw=0.5; \atomtranh=0.6; \next= \atomtranw*1.2; \brratio=0.7; \photonlinescl=0.8; }
 
This self-regulating mechanism is a fundamental and intriguing 
aspect of the MPMA process.
 As a result,
 ${W_{a_o}}$ cannot be increased arbitrarily but likely
 tends to exhibit a near-saturation behavior below a scale 
 of $10^{9}$ s$^{-1}$ or so. 
 Unlike other saturation behaviors in the field of 
 matter-light interaction, the near-saturation of 
 ${W_{a_o}}$ is governed by the principle of relativistic causality.
 This distinctive feature highlights the profound interplay between 
 quantum mechanics and the principles of relativity, offering 
 a glimpse into their interconnected nature.

A fundamental application of the MPMA process is to 
enhance ultraweak atomic absorption transitions to observable levels.
For instance, consider an  electric multipole absorption,
where the absorbed electric $2^{j}$-pole ($Ej$) photon \cite{berestetskiiQEDbook}
 carries an angular momentum of $j\hbar$ ($j=1,2,3,\dots$). 
 General analysis indicates that transition probability  for the $Ej$ transition  scales 
 approximately as $(a/\lambda)^{2j}$ \cite{berestetskiiQEDbook}, where $a$ is the linear 
 size of the atom and $\lambda$ is the wavelength of the photon involved 
 \cite{berestetskiiQEDbook}.  
Typically, $(a/\lambda) \approx 10^{-4}$, so the probability for the $E2$ transition 
is on the order of $10^{-8}$ relative to the electric dipole 
($E1$) transition.   Similarly, the probability for the 
 $E3$ transition is  around $10^{-16}$, 
the $E4$ transition is around $10^{-24}$, and so on.
Accordingly, the $E4$ transition rate is estimated to   
be around $10^{-16}$ s$^{-1}$ \cite{e4rate}, making it extremely 
challenging to observe.
 Higher multipole transitions,
such as $E5$ and $E6$, are generally undetectable in their unenhanced form.
However, the MPMA process can be used to elevate these ultraweak transition 
rates to levels comparable to $E1$ transitions,
 enabling experimental investigation of these otherwise elusive transitions. 

\begin{figure}
\captionsetup{width=.92\linewidth}
\begin{tikzpicture}
\begin{scope}[xshift=0, scale=1.0]

\myphotonlineRatioThick.apply(\wx -3.5*\linexinc, -3*\wint -3.5*\lineyinc, \longline*\wsz, -2*\wint+\longline*\wsz, \asz, webblue, \large)
\myphotonlineRatioThick.apply(\wx -2.5*\linexinc, -3*\wint -2.5*\lineyinc, \longline*\wsz, -2*\wint+\longline*\wsz, \asz, webblue, \large)
\myphotonlineRatioThick.apply(\wx -1.5*\linexinc, -3*\wint -1.5*\lineyinc, \longline*\wsz, -2*\wint+\longline*\wsz, \asz, webblue, \large)
\myphotonlineRatio.apply(\wx, -3*\wint, \longline*\wsz, -2*\wint+ \longline*\wsz, \asz, redgold, 1.4*\large)
\myphotonlineRatio.apply(\wx + \linexinc, -3*\wint + \lineyinc, \longline*\wsz, -2*\wint+\longline*\wsz, \asz, redgold, 1.4*\large)
\myphotonlineRatio.apply(\wx + 2*\linexinc, -3*\wint + 2*\lineyinc, \longline*\wsz, -2*\wint+\longline*\wsz, \asz, redgold, 1.4*\large)
\myphotonlineRatio.apply(\wx + 3*\linexinc, -3*\wint + 3*\lineyinc, \longline*\wsz, -2*\wint+\longline*\wsz, \asz, redgold, 1.4*\large)
\myphotonlineRatio.apply(\wx + 4*\linexinc, -3*\wint + 4*\lineyinc, \longline*\wsz, -2*\wint+\longline*\wsz, \asz, redgold, 1.4*\large)
\myphotonlineRatio.apply(\wx + 5*\linexinc, -3*\wint + 5*\lineyinc, \longline*\wsz, -2*\wint+\longline*\wsz, \asz, redgold, 1.4*\large)

\node[circle, shading=ball, ball color=unexcited-rcol,  minimum size=2pt, scale=1.25*\atomsize,
text=white] (ball) at(\atAx,  \atAy){} ;
\node[circle, shading=ball, ball color=unexcited-rcol,  minimum size=2pt, scale=1.*\atomsize,
text=white] (ball) at(\atBx,  \atBy){} ;
\node[circle, shading=ball, ball color=unexcited-rcol,  minimum size=2pt, scale=1.*\atomsize,
text=white] (ball) at(\atCx,  \atCy){} ;
\node[circle, shading=ball, ball color=unexcited-rcol,  minimum size=2pt, scale=1.*\atomsize,
text=white] (ball) at(\atDx,  \atDy){} ;

\myatomTransition.two_levels(\atAlevelx, \atAlevely, 0.3*\atomtranw, 0.3*\atomtranh, 0, gray2, bcol)
\draw[dash pattern={on \padint off \padint on \padint off \padint on \padint}](\atAx, \atAy-0.3*\pad)--(\atAlevelx-0.3*\pad ,\atAlevely +0.3*\pad );
\draw[dash pattern={on \padint off \padint on \padint off \padint on \padint}](\atAx, \atAy+ 0.3*\pad)--(\atAlevelx-0.3*\pad ,\atAlevely + 0.3*\atomtranh + 0.3*\pad );
\draw[dashed,dash pattern=on \padint off \padint, gray2, -stealth](\atAlevelx+ 0.5*0.3*\atomtranw, \atAlevely )--(\atAlevelx+ 0.5*0.3*\atomtranw, \atAlevely +  0.3*\atomtranh);

\draw[dashed,dash pattern=on \padint off \padint, gray2](\atAlevelx+ 0.5*0.3* \atomtranw, \atAlevely +  0.5*0.3*\atomtranh) ellipse ( 0.25*\atomtranw cm and 0.25*\atomtranh cm);
\draw[thick,gray2](\atAlevelx+ 0.5*0.3 *\atomtranw, \atAlevely ) circle (0.05);

\myatomTransition.two_levels(\atDlevelx, \atDlevely, 0.3*\atomtranw, \rbfac * 0.3*\atomtranh, 0, gray2, rcol)
\draw[dash pattern={on \padint off \padint on \padint off \padint on \padint}](\atDx, \atDy-0.0*\pad)--(\atDlevelx-0.3*\pad ,\atDlevely - 0.3*\pad );
\draw[dash pattern={on \padint off \padint on \padint off \padint on \padint}](\atDx, \atDy+ 0.8*\pad)--(\atDlevelx-0.3*\pad ,\atDlevely + 0.3*\rbfac*\atomtranh - 0.3*\pad );
\draw[dashed,dash pattern=on \padint off \padint, gray2, -stealth](\atDlevelx+ 0.5*0.3* \atomtranw, \atDlevely )--(\atDlevelx+ 0.5*0.3*\atomtranw, \atDlevely +  0.3*\rbfac*\atomtranh);
\draw[dashed,dash pattern=on \padint off \padint, gray2](\atDlevelx+ 0.5*0.3* \atomtranw, \atDlevely +  \rbfac*0.5*0.3*\atomtranh) ellipse ( 0.25*\atomtranw cm and 0.25*\atomtranh *\rbfac*1.3 cm);
\draw[thick,gray2](\atDlevelx+ 0.5*0.3* \atomtranw, \atDlevely ) circle (0.05);

\end{scope}
\begin{scope}[xshift=100, scale=1.0]
\tikzmath{ \lineyinc= \lineyinc * 1.2;}
\myphotonlineRatioThick.apply(\wx -1.2*\linexinc+\lineshift, -3*\wint -1.2*\lineyinc, \longline*\wsz, -2*\wint+\longline*\wsz, \asz, webblue, \large)
\myphotonlineRatioThick.apply(\wx -2.2*\linexinc+\lineshift, -3*\wint -2.2*\lineyinc, \longline*\wsz, -2*\wint+\longline*\wsz, \asz, webblue, \large)

\myphotonlineRatio.apply(\wx +\lineshift, -3*\wint, \longline*\wsz, -2*\wint+ \longline*\wsz, \asz, redgold, 1.4*\large)
\myphotonlineRatio.apply(\wx + 1.0*\linexinc +\lineshift, -3*\wint + 1.0*\lineyinc, \longline*\wsz, -2*\wint+\longline*\wsz, \asz, redgold, 1.4*\large)
\myphotonlineRatio.apply(\wx + 2.0*\linexinc +\lineshift, -3*\wint + 2.0*\lineyinc, \longline*\wsz, -2*\wint+\longline*\wsz, \asz, redgold, 1.4*\large)

\node[circle, shading=ball, ball color=bcol,  minimum size=2pt, scale=1.25*\atomsize,
text=white] (ball) at(\atAx,  \atAy){} ;
\node[circle, shading=ball, ball color=rcol,  minimum size=2pt, scale=1.*\atomsize,
text=white] (ball) at(\atBx,  \atBy){} ;
\node[circle, shading=ball, ball color=rcol,  minimum size=2pt, scale=1.*\atomsize,
text=white] (ball) at(\atCx,  \atCy){} ;
\node[circle, shading=ball, ball color= rcol, minimum size=2pt, scale=1.*\atomsize,
text=white] (ball) at(\atDx,  \atDy){} ;
\myatomTransition.two_levels(\atAlevelx, \atAlevely, 0.3*\atomtranw, 0.3*\atomtranh, 0, gray2, bcol)
\draw[dash pattern={on \padint off \padint on \padint off \padint on \padint}](\atAx, \atAy-0.3*\pad)--(\atAlevelx- 0.3*\pad ,\atAlevely +-.3*\pad );
\draw[dash pattern={on \padint off \padint on \padint off \padint on \padint}](\atAx, \atAy+ 0.3*\pad)--(\atAlevelx-0.3*\pad ,\atAlevely + 0.3*\atomtranh + 0.3*\pad );
\draw[dashed,dash pattern=on \padint off \padint, gray2](\atAlevelx+ 0.5*0.3* \atomtranw, \atAlevely +  0.5*0.3*\atomtranh) ellipse ( 0.25*\atomtranw cm and 0.25*\atomtranh cm);
\draw[thick,blue](\atAlevelx+ 0.5*0.3* \atomtranw, \atAlevely+ 0.3*\atomtranh ) circle (0.05);

\myatomTransition.two_levels(\atDlevelx, \atDlevely, 0.3*\atomtranw, \rbfac * 0.3*\atomtranh, 0, gray2, rcol)
\draw[dash pattern={on \padint off \padint on \padint off \padint on \padint}](\atDx, \atDy +0.0*\pad)--(\atDlevelx-0.3*\pad ,\atDlevely - 0.3*\pad );
\draw[dash pattern={on \padint off \padint on \padint off \padint on \padint}](\atDx, \atDy+ 0.8*\pad)--(\atDlevelx-0.3*\pad ,\atDlevely + 0.3*\rbfac*\atomtranh - 0.3*\pad );
\draw[dashed,dash pattern=on \padint off \padint, gray2](\atDlevelx+ 0.5*0.3* \atomtranw, \atDlevely +  \rbfac*0.5*0.3*\atomtranh) ellipse ( 0.25*\atomtranw cm and 0.25*\atomtranh *\rbfac*1.3 cm);
\draw[thick,rcol](\atDlevelx+ 0.5*0.3* \atomtranw, \atDlevely+\rbfac*0.3*\atomtranh ) circle (0.05);

\end{scope}
\end{tikzpicture}
\caption{ \justifying Schematic illustration of an $Ej$-photon-absorption  process 
involving a  simultaneous joint excitation of a four-atom system. The $A^\prime$-species atom 
(the large ball) absorbs an $Ej$ photon (blue lines), while
 each of  three $B$-species atoms absorbs a laser photon (red  lines). }
 \label{fig:Ej_mpma} 
\end{figure}
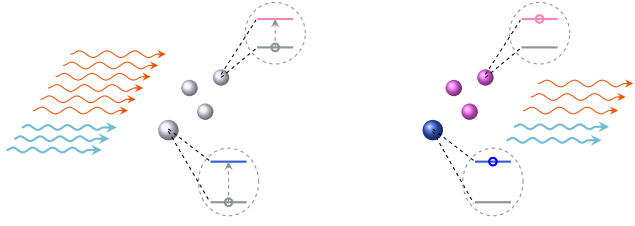

The integration of the MPMA process into an  atomic $Ej$-photon
absorption 
transition of a certain species of atoms, denoted as $A^\prime$-species atoms, 
can generally be accomplished  by adjusting the frequencies 
of the stimulating (laser) photons to satisfy 
 energy conservation in a joint transition
 process involving a corresponding $m$-atom system. In this process, 
 one $A^\prime$-species atom absorbs an $Ej$-photon, 
 while simultaneously, $m-1$
$B$-species atoms are excited by absorbing 
$m-1$ laser photons (see Fig. \ref{fig:Ej_mpma} for an example with $m=4$).
 The  laser frequency $\Omega_{\mathfrak{L}_2}$ is close to, but
not equal to,  $\omega_b$, and the energy conservation is expressed as:
\begin{equation}
\hbar \Omega_{Ej} + (m-1) \hbar \Omega_{\mathfrak{L}_2}= \hbar \omega_{a,Ej} + (m-1) \hbar \omega_b ,
\end{equation}
where $\Omega_{Ej}$ is the frequency of the $Ej$ photon,and
$\omega_{a,Ej}$ denotes the $Ej$-transition frequency of an $A^\prime$-species
atom. 
Although the transition rate for this MPMA process within a single $m$-atom 
system is extraordinarily small—partly due to the extremely small strength
of the $Ej$ transition in the $A^\prime$-species atom—the overall transition
rate for a given $A^\prime$-species atom in a two-species atomic gas can 
become significant. This overall rate benefits from an enhancement 
factor of $N_{b\mathfrak{o}}^{m-1}$, which can surpass many orders 
of magnitude, rendering the rate substantial enough to reach an observable level.
 
 \tikzmath{\atomtranx=0.0; \atomtrany= 0.0; \atomtranw=1.6; \atomtranh=1.8; }

       \tikzmath{ \atAy= -16*\wint; \atBx= \atAx + 0.4*\atgap;
      \atBy= \atAy + 0.8*\atgap; \atCx= \atAx + 0.7*\atgap; \atCy= \atAy + 0.35*\atgap;
      \atDx= \atAx + 1.0*\atgap; \atDy= \atAy + 1.0*\atgap;  \atDlevelx= \atDx +0.6*\levelgap;
       \atDlevely= \atDy + 0.5*\levelgap;  \atAlevelx= \atAx +0.7*\levelgap;
       \atAlevely= \atAy - 1.2*\levelgap;}

A flux of $Ej$ photons is essential for experimentally investigating
 these enhanced atomic absorption transitions. Technologically, twisted 
 light \cite{Allen1999, Yao2011,Zannotti2016, Russell2017, Babiker2019}, such as 
 Laguerre–Gaussian and Bessel beams, offers a feasible and natural
  solution, as it delivers photons carrying high angular momentum.
  The interaction between twisted photons and atoms has been 
  theoretically studied \cite{Andrei2013atomTwisted, Marggraft2014wistedatomstheor2, SchulzE3transition}, but 
  experimental observation remains an open challenge. One major difficulty arises 
  from the inherently low transition rates of multipole transitions, as discussed above. 
  Recently, a quadrupole transition in a trapped $^{40}$Ca$^{+}$ 
  ion has been observed via the absorption of a twisted photon
   \cite{Schmiegelow2016ionabsorbption}. With MPMA enhancement, atomic absorption 
   of twisted photons could become significantly more detectable.
   Furthermore, it would be interesting to explore whether MPMA enhancement 
   can also be utilized to facilitate the emission of 
   high-multipole photons from atomic gases.

In summary,  MPMA processes represent a fundamental 
and fascinating aspect of matter-light interactions, exhibiting 
unusual behaviors such as the near-saturation of the atomic 
transition rates under certain circumstances. MPMA processes can 
be used to
 amplify ultraweak atomic transitions by many orders  
 of magnitude,  offering a powerful tool for probing atomic 
 phenomena that would otherwise remain inaccessible 
 to experimental observation. 
 The super-enhancement mechanism underlying MPMA processes is deeply 
 rooted in the fundamental principles of quantum mechanics, 
 holding substantial potential to enrich our understanding 
 of light-matter interactions and advance our comprehension 
 of quantum collective phenomena.


\bibliographystyle{apsrev4-1}

\bibliography{laserbib}

\begin{thebibliography}{26}%
\makeatletter
\providecommand \@ifxundefined [1]{%
 \@ifx{#1\undefined}
}%
\providecommand \@ifnum [1]{%
 \ifnum #1\expandafter \@firstoftwo
 \else \expandafter \@secondoftwo
 \fi
}%
\providecommand \@ifx [1]{%
 \ifx #1\expandafter \@firstoftwo
 \else \expandafter \@secondoftwo
 \fi
}%
\providecommand \natexlab [1]{#1}%
\providecommand \enquote  [1]{``#1''}%
\providecommand \bibnamefont  [1]{#1}%
\providecommand \bibfnamefont [1]{#1}%
\providecommand \citenamefont [1]{#1}%
\providecommand \href@noop [0]{\@secondoftwo}%
\providecommand \href [0]{\begingroup \@sanitize@url \@href}%
\providecommand \@href[1]{\@@startlink{#1}\@@href}%
\providecommand \@@href[1]{\endgroup#1\@@endlink}%
\providecommand \@sanitize@url [0]{\catcode `\\12\catcode `\$12\catcode
  `\&12\catcode `\#12\catcode `\^12\catcode `\_12\catcode `\%12\relax}%
\providecommand \@@startlink[1]{}%
\providecommand \@@endlink[0]{}%
\providecommand \url  [0]{\begingroup\@sanitize@url \@url }%
\providecommand \@url [1]{\endgroup\@href {#1}{\urlprefix }}%
\providecommand \urlprefix  [0]{URL }%
\providecommand \Eprint [0]{\href }%
\providecommand \doibase [0]{http://dx.doi.org/}%
\providecommand \selectlanguage [0]{\@gobble}%
\providecommand \bibinfo  [0]{\@secondoftwo}%
\providecommand \bibfield  [0]{\@secondoftwo}%
\providecommand \translation [1]{[#1]}%
\providecommand \BibitemOpen [0]{}%
\providecommand \bibitemStop [0]{}%
\providecommand \bibitemNoStop [0]{.\EOS\space}%
\providecommand \EOS [0]{\spacefactor3000\relax}%
\providecommand \BibitemShut  [1]{\csname bibitem#1\endcsname}%
\let\auto@bib@innerbib\@empty
\bibitem [{\citenamefont {Cowan}\ \emph {et~al.}(1956)\citenamefont {Cowan},
  \citenamefont {Reines}, \citenamefont {Harrison}, \citenamefont {Kruse},\
  and\ \citenamefont {McGuire}}]{neutrino}%
  \BibitemOpen
  \bibfield  {author} {\bibinfo {author} {\bibfnamefont {C.~L.}\ \bibnamefont
  {Cowan}}, \bibinfo {author} {\bibfnamefont {F.}~\bibnamefont {Reines}},
  \bibinfo {author} {\bibfnamefont {F.~B.}\ \bibnamefont {Harrison}}, \bibinfo
  {author} {\bibfnamefont {H.~W.}\ \bibnamefont {Kruse}}, \ and\ \bibinfo
  {author} {\bibfnamefont {A.~D.}\ \bibnamefont {McGuire}},\ }\href@noop {}
  {\bibfield  {journal} {\bibinfo  {journal} {Science}\ }\textbf {\bibinfo
  {volume} {124}},\ \bibinfo {pages} {103} (\bibinfo {year}
  {1956})}\BibitemShut {NoStop}%
\bibitem [{\citenamefont {Kaiser}\ and\ \citenamefont
  {Garrett}(1961)}]{twophoton}%
  \BibitemOpen
  \bibfield  {author} {\bibinfo {author} {\bibfnamefont {W.}~\bibnamefont
  {Kaiser}}\ and\ \bibinfo {author} {\bibfnamefont {C.~G.~B.}\ \bibnamefont
  {Garrett}},\ }\href@noop {} {\bibfield  {journal} {\bibinfo  {journal} {Phys.
  Rev. Lett.}\ }\textbf {\bibinfo {volume} {7}},\ \bibinfo {pages} {229}
  (\bibinfo {year} {1961})}\BibitemShut {NoStop}%
\bibitem [{\citenamefont {White}(1981)}]{WhiteAtomgasTlBa}%
  \BibitemOpen
  \bibfield  {author} {\bibinfo {author} {\bibfnamefont {J.~C.}\ \bibnamefont
  {White}},\ }\href {\doibase 10.1364/OL.6.000242} {\bibfield  {journal}
  {\bibinfo  {journal} {Optics Letters}\ }\textbf {\bibinfo {volume} {6}},\
  \bibinfo {pages} {242} (\bibinfo {year} {1981})}\BibitemShut {NoStop}%
\bibitem [{\citenamefont {Pedrozo-Pe\~nafiel}\ \emph
  {et~al.}(2012)\citenamefont {Pedrozo-Pe\~nafiel}, \citenamefont {Paiva},
  \citenamefont {Vivanco}, \citenamefont {Bagnato},\ and\ \citenamefont
  {Farias}}]{NaAtoms}%
  \BibitemOpen
  \bibfield  {author} {\bibinfo {author} {\bibfnamefont {E.}~\bibnamefont
  {Pedrozo-Pe\~nafiel}}, \bibinfo {author} {\bibfnamefont {R.~R.}\ \bibnamefont
  {Paiva}}, \bibinfo {author} {\bibfnamefont {F.~J.}\ \bibnamefont {Vivanco}},
  \bibinfo {author} {\bibfnamefont {V.~S.}\ \bibnamefont {Bagnato}}, \ and\
  \bibinfo {author} {\bibfnamefont {K.~M.}\ \bibnamefont {Farias}},\ }\href
  {\doibase 10.1103/PhysRevLett.108.253004} {\bibfield  {journal} {\bibinfo
  {journal} {Phys. Rev. Lett.}\ }\textbf {\bibinfo {volume} {108}},\ \bibinfo
  {pages} {253004} (\bibinfo {year} {2012})}\BibitemShut {NoStop}%
\bibitem [{\citenamefont {Leite}\ and\ \citenamefont
  {Araujo}(1980)}]{rios1980lineshape}%
  \BibitemOpen
  \bibfield  {author} {\bibinfo {author} {\bibfnamefont {J.~R.}\ \bibnamefont
  {Leite}}\ and\ \bibinfo {author} {\bibfnamefont {C.~B.~D.}\ \bibnamefont
  {Araujo}},\ }\href {\doibase 10.1016/0009-2614(80)85514-3} {\bibfield
  {journal} {\bibinfo  {journal} {Chemical Physics Letters}\ }\textbf {\bibinfo
  {volume} {73}},\ \bibinfo {pages} {71} (\bibinfo {year} {1980})}\BibitemShut
  {NoStop}%
\bibitem [{\citenamefont {Andrews}\ and\ \citenamefont
  {Harlow}(1983)}]{andrews1983cooperative}%
  \BibitemOpen
  \bibfield  {author} {\bibinfo {author} {\bibfnamefont {D.~L.}\ \bibnamefont
  {Andrews}}\ and\ \bibinfo {author} {\bibfnamefont {M.}~\bibnamefont
  {Harlow}},\ }\href@noop {} {\bibfield  {journal} {\bibinfo  {journal} {The
  Journal of Chemical Physics}\ }\textbf {\bibinfo {volume} {78}},\ \bibinfo
  {pages} {1088} (\bibinfo {year} {1983})}\BibitemShut {NoStop}%
\bibitem [{\citenamefont {Nayfeh}\ and\ \citenamefont
  {Hillard}(1984)}]{Nayfeh1984}%
  \BibitemOpen
  \bibfield  {author} {\bibinfo {author} {\bibfnamefont {M.~H.}\ \bibnamefont
  {Nayfeh}}\ and\ \bibinfo {author} {\bibfnamefont {G.~B.}\ \bibnamefont
  {Hillard}},\ }\href {\doibase 10.1103/PhysRevA.29.1907} {\bibfield  {journal}
  {\bibinfo  {journal} {Physical Review A}\ }\textbf {\bibinfo {volume} {29}},\
  \bibinfo {pages} {1907} (\bibinfo {year} {1984})}\BibitemShut {NoStop}%
\bibitem [{\citenamefont {Kim}\ and\ \citenamefont {Agarwal}(1998)}]{Kim1998}%
  \BibitemOpen
  \bibfield  {author} {\bibinfo {author} {\bibfnamefont {M.~S.}\ \bibnamefont
  {Kim}}\ and\ \bibinfo {author} {\bibfnamefont {G.~S.}\ \bibnamefont
  {Agarwal}},\ }\href {\doibase 10.1103/PhysRevA.57.3059} {\bibfield  {journal}
  {\bibinfo  {journal} {Physical Review A}\ }\textbf {\bibinfo {volume} {57}},\
  \bibinfo {pages} {3059} (\bibinfo {year} {1998})}\BibitemShut {NoStop}%
\bibitem [{\citenamefont {Muthukrishnan}\ \emph {et~al.}(2004)\citenamefont
  {Muthukrishnan}, \citenamefont {Agarwal},\ and\ \citenamefont
  {Scully}}]{Muthukrishnan2004}%
  \BibitemOpen
  \bibfield  {author} {\bibinfo {author} {\bibfnamefont {A.}~\bibnamefont
  {Muthukrishnan}}, \bibinfo {author} {\bibfnamefont {G.~S.}\ \bibnamefont
  {Agarwal}}, \ and\ \bibinfo {author} {\bibfnamefont {M.~O.}\ \bibnamefont
  {Scully}},\ }\href {\doibase 10.1103/PhysRevLett.93.093002} {\bibfield
  {journal} {\bibinfo  {journal} {Physical Review Letters}\ }\textbf {\bibinfo
  {volume} {93}},\ \bibinfo {pages} {093002} (\bibinfo {year}
  {2004})}\BibitemShut {NoStop}%
\bibitem [{\citenamefont {Zheng}\ \emph {et~al.}(2013)\citenamefont {Zheng},
  \citenamefont {Saldanha}, \citenamefont {Leite},\ and\ \citenamefont
  {Fabre}}]{Zheng2013}%
  \BibitemOpen
  \bibfield  {author} {\bibinfo {author} {\bibfnamefont {Z.}~\bibnamefont
  {Zheng}}, \bibinfo {author} {\bibfnamefont {P.~L.}\ \bibnamefont {Saldanha}},
  \bibinfo {author} {\bibfnamefont {J.~R.~R.}\ \bibnamefont {Leite}}, \ and\
  \bibinfo {author} {\bibfnamefont {C.}~\bibnamefont {Fabre}},\ }\href
  {\doibase 10.1103/PhysRevA.88.033822} {\bibfield  {journal} {\bibinfo
  {journal} {Physical Review A}\ }\textbf {\bibinfo {volume} {88}},\ \bibinfo
  {pages} {033822} (\bibinfo {year} {2013})}\BibitemShut {NoStop}%
\bibitem [{\citenamefont {Hettich}\ \emph {et~al.}(2002)\citenamefont
  {Hettich}, \citenamefont {Schmitt}, \citenamefont {Zitzmann}, \citenamefont
  {K{\"u}hn}, \citenamefont {Gerhardt},\ and\ \citenamefont
  {Sandoghdar}}]{twoMolecules}%
  \BibitemOpen
  \bibfield  {author} {\bibinfo {author} {\bibfnamefont {C.}~\bibnamefont
  {Hettich}}, \bibinfo {author} {\bibfnamefont {C.}~\bibnamefont {Schmitt}},
  \bibinfo {author} {\bibfnamefont {J.}~\bibnamefont {Zitzmann}}, \bibinfo
  {author} {\bibfnamefont {S.}~\bibnamefont {K{\"u}hn}}, \bibinfo {author}
  {\bibfnamefont {I.}~\bibnamefont {Gerhardt}}, \ and\ \bibinfo {author}
  {\bibfnamefont {V.}~\bibnamefont {Sandoghdar}},\ }\href {\doibase
  10.1126/science.1075606} {\bibfield  {journal} {\bibinfo  {journal}
  {Science}\ }\textbf {\bibinfo {volume} {298}},\ \bibinfo {pages} {385}
  (\bibinfo {year} {2002})}\BibitemShut {NoStop}%
\bibitem [{\citenamefont {Yu}(2025)}]{yu1}%
  \BibitemOpen
  \bibfield  {author} {\bibinfo {author} {\bibfnamefont {Y.}~\bibnamefont
  {Yu}},\ }\href {https://arxiv.org/abs/2504.05773} {\bibfield  {journal}
  {\bibinfo  {journal} {arXiv:2504.05773}\ } (\bibinfo {year}
  {2025})}\BibitemShut {NoStop}%
\bibitem [{den()}]{densityStates}%
  \BibitemOpen
  \href@noop {} {}\bibinfo {note} {Formally, $\rho(E_f) = \delta(E_f -
  \varepsilon_e^a - (m-1)\varepsilon_e^b)$. However, due to the intrinsic
  coupling between the excited atoms and the quantum electromagnetic field, the
  excited levels of each atom exhibit a resonant-state nature and possess a
  finite width (see, {\it e.g.}, \cite{Tannoudji}). As a result, the $m$-atom
  system also acquires a finite width at $E_f = \varepsilon_e^a -
  (m-1)\varepsilon_e^b $, and the density of states could be approximated as:
  $\rho(E_f) \sim \frac{\Gamma_{eff}/2}{(E_f - \varepsilon_e^a -
  (m-1)\varepsilon_e^b)^2 + (\Gamma_{eff}/2)^2}$, where $\Gamma_{eff}$ is
  comparable to the maximum natural widths of the quantum states $|e_a\rangle$
  and $|g_{b_i}\rangle$, differing by a factor between unity and $m$. Moreover,
  thermal broadening of the excited states can be incorporated into $\rho(E_f)$
  as well.}\BibitemShut {Stop}%
\bibitem [{f_a()}]{f_absorb}%
  \BibitemOpen
  \href@noop {} {}\bibinfo {note} {The term $f^2_o(m)$ in the suppression
  factor associated with quantum interference has been incorporated into
  $[f(m)]^{2m} $ in Eq. {\ref{eq:w2}}.}\BibitemShut {Stop}%
\bibitem [{\citenamefont {V.~B.~Berestetskii}\ and\ \citenamefont
  {Pitaevskii}(2008)}]{berestetskiiQEDbook}%
  \BibitemOpen
  \bibfield  {author} {\bibinfo {author} {\bibfnamefont {E.~M.~L.}\
  \bibnamefont {V.~B.~Berestetskii}}\ and\ \bibinfo {author} {\bibfnamefont
  {L.~P.}\ \bibnamefont {Pitaevskii}},\ }\href@noop {} {\emph {\bibinfo {title}
  {Quantum Electrodynamics}}},\ \bibinfo {edition} {2nd}\ ed.\ (\bibinfo
  {publisher} {Elsevier (Singapore) Pte Ltd.},\ \bibinfo {address}
  {Singapore},\ \bibinfo {year} {2008})\BibitemShut {NoStop}%
\bibitem [{e4r()}]{e4rate}%
  \BibitemOpen
  \href@noop {} {}\bibinfo {note} {The rate here can refer to either the
  spontaneous $E4$-photon emission rate or the $E4$-photon absorption rate.
  When it refers to the $E4$-photon absorption rate of $10^{-16}$ s$^{-1}$, it
  should be noted that the requisite $E4$-photon field possesses a certain
  strength, corresponding to a photon density where the number of photons
  within a volume of $\lambda^3$—where $\lambda$ is the wavelength of the
  $E4$ photon—is approximately on the order of $10^{-7}$.}\BibitemShut
  {Stop}%
\bibitem [{\citenamefont {Allen}\ \emph {et~al.}(1999)\citenamefont {Allen},
  \citenamefont {Padgett},\ and\ \citenamefont {Babiker}}]{Allen1999}%
  \BibitemOpen
  \bibfield  {author} {\bibinfo {author} {\bibfnamefont {L.}~\bibnamefont
  {Allen}}, \bibinfo {author} {\bibfnamefont {M.~J.}\ \bibnamefont {Padgett}},
  \ and\ \bibinfo {author} {\bibfnamefont {M.}~\bibnamefont {Babiker}},\ }in\
  \href {\doibase 10.1016/S0079-6638(08)70391-3} {\emph {\bibinfo {booktitle}
  {Progress in Optics}}},\ Vol.~\bibinfo {volume} {39},\ \bibinfo {editor}
  {edited by\ \bibinfo {editor} {\bibfnamefont {E.}~\bibnamefont {Wolf}}}\
  (\bibinfo  {publisher} {Elsevier},\ \bibinfo {year} {1999})\ pp.\ \bibinfo
  {pages} {291--372}\BibitemShut {NoStop}%
\bibitem [{\citenamefont {Yao}\ and\ \citenamefont {Padgett}(2011)}]{Yao2011}%
  \BibitemOpen
  \bibfield  {author} {\bibinfo {author} {\bibfnamefont {A.~M.}\ \bibnamefont
  {Yao}}\ and\ \bibinfo {author} {\bibfnamefont {M.~J.}\ \bibnamefont
  {Padgett}},\ }\href {\doibase 10.1364/AOP.3.000161} {\bibfield  {journal}
  {\bibinfo  {journal} {Advances in Optics and Photonics}\ }\textbf {\bibinfo
  {volume} {3}},\ \bibinfo {pages} {161} (\bibinfo {year} {2011})}\BibitemShut
  {NoStop}%
\bibitem [{\citenamefont {Zannotti}\ \emph {et~al.}(2016)\citenamefont
  {Zannotti}, \citenamefont {Diebel}, \citenamefont {Boguslawski},\ and\
  \citenamefont {Denz}}]{Zannotti2016}%
  \BibitemOpen
  \bibfield  {author} {\bibinfo {author} {\bibfnamefont {A.}~\bibnamefont
  {Zannotti}}, \bibinfo {author} {\bibfnamefont {F.}~\bibnamefont {Diebel}},
  \bibinfo {author} {\bibfnamefont {M.}~\bibnamefont {Boguslawski}}, \ and\
  \bibinfo {author} {\bibfnamefont {C.}~\bibnamefont {Denz}},\ }\href {\doibase
  10.1002/adom.201600629} {\bibfield  {journal} {\bibinfo  {journal} {Advanced
  Optical Materials}\ }\textbf {\bibinfo {volume} {5}},\ \bibinfo {pages}
  {1600629} (\bibinfo {year} {2016})}\BibitemShut {NoStop}%
\bibitem [{\citenamefont {Russell}\ \emph {et~al.}(2017)\citenamefont
  {Russell}, \citenamefont {Beravat},\ and\ \citenamefont
  {Wong}}]{Russell2017}%
  \BibitemOpen
  \bibfield  {author} {\bibinfo {author} {\bibfnamefont {P.~S.~J.}\
  \bibnamefont {Russell}}, \bibinfo {author} {\bibfnamefont {R.}~\bibnamefont
  {Beravat}}, \ and\ \bibinfo {author} {\bibfnamefont {G.~K.~L.}\ \bibnamefont
  {Wong}},\ }\href {\doibase 10.1098/rsta.2015.0440} {\bibfield  {journal}
  {\bibinfo  {journal} {Philosophical Transactions of the Royal Society A}\
  }\textbf {\bibinfo {volume} {375}},\ \bibinfo {pages} {20150440} (\bibinfo
  {year} {2017})}\BibitemShut {NoStop}%
\bibitem [{\citenamefont {Babiker}\ \emph {et~al.}(2019)\citenamefont
  {Babiker}, \citenamefont {Andrews},\ and\ \citenamefont
  {Lembessis}}]{Babiker2019}%
  \BibitemOpen
  \bibfield  {author} {\bibinfo {author} {\bibfnamefont {M.}~\bibnamefont
  {Babiker}}, \bibinfo {author} {\bibfnamefont {D.~L.}\ \bibnamefont
  {Andrews}}, \ and\ \bibinfo {author} {\bibfnamefont {V.~E.}\ \bibnamefont
  {Lembessis}},\ }\href {\doibase 10.1088/2040-8986/aaf6d3} {\bibfield
  {journal} {\bibinfo  {journal} {Journal of Optics}\ }\textbf {\bibinfo
  {volume} {21}},\ \bibinfo {pages} {013001} (\bibinfo {year}
  {2019})}\BibitemShut {NoStop}%
\bibitem [{\citenamefont {Afanasev}\ \emph {et~al.}(2013)\citenamefont
  {Afanasev}, \citenamefont {Carlson},\ and\ \citenamefont
  {Mukherjee}}]{Andrei2013atomTwisted}%
  \BibitemOpen
  \bibfield  {author} {\bibinfo {author} {\bibfnamefont {A.}~\bibnamefont
  {Afanasev}}, \bibinfo {author} {\bibfnamefont {C.~E.}\ \bibnamefont
  {Carlson}}, \ and\ \bibinfo {author} {\bibfnamefont {A.}~\bibnamefont
  {Mukherjee}},\ }\href {\doibase 10.1103/PhysRevA.88.033841} {\bibfield
  {journal} {\bibinfo  {journal} {Phys. Rev. A}\ }\textbf {\bibinfo {volume}
  {88}},\ \bibinfo {pages} {033841} (\bibinfo {year} {2013})}\BibitemShut
  {NoStop}%
\bibitem [{\citenamefont {Scholz-Marggraf}\ \emph {et~al.}(2014)\citenamefont
  {Scholz-Marggraf}, \citenamefont {Fritzsche}, \citenamefont {Serbo},
  \citenamefont {Afanasev},\ and\ \citenamefont
  {Surzhykov}}]{Marggraft2014wistedatomstheor2}%
  \BibitemOpen
  \bibfield  {author} {\bibinfo {author} {\bibfnamefont {H.~M.}\ \bibnamefont
  {Scholz-Marggraf}}, \bibinfo {author} {\bibfnamefont {S.}~\bibnamefont
  {Fritzsche}}, \bibinfo {author} {\bibfnamefont {V.~G.}\ \bibnamefont
  {Serbo}}, \bibinfo {author} {\bibfnamefont {A.}~\bibnamefont {Afanasev}}, \
  and\ \bibinfo {author} {\bibfnamefont {A.}~\bibnamefont {Surzhykov}},\ }\href
  {\doibase 10.1103/PhysRevA.90.013425} {\bibfield  {journal} {\bibinfo
  {journal} {Phys. Rev. A}\ }\textbf {\bibinfo {volume} {90}},\ \bibinfo
  {pages} {013425} (\bibinfo {year} {2014})}\BibitemShut {NoStop}%
\bibitem [{\citenamefont {Schulz}\ \emph {et~al.}(2020)\citenamefont {Schulz},
  \citenamefont {Peshkov}, \citenamefont {M\"uller}, \citenamefont {Lange},
  \citenamefont {Huntemann}, \citenamefont {Tamm}, \citenamefont {Peik},\ and\
  \citenamefont {Surzhykov}}]{SchulzE3transition}%
  \BibitemOpen
  \bibfield  {author} {\bibinfo {author} {\bibfnamefont {S.~A.-L.}\
  \bibnamefont {Schulz}}, \bibinfo {author} {\bibfnamefont {A.~A.}\
  \bibnamefont {Peshkov}}, \bibinfo {author} {\bibfnamefont {R.~A.}\
  \bibnamefont {M\"uller}}, \bibinfo {author} {\bibfnamefont {R.}~\bibnamefont
  {Lange}}, \bibinfo {author} {\bibfnamefont {N.}~\bibnamefont {Huntemann}},
  \bibinfo {author} {\bibfnamefont {C.}~\bibnamefont {Tamm}}, \bibinfo {author}
  {\bibfnamefont {E.}~\bibnamefont {Peik}}, \ and\ \bibinfo {author}
  {\bibfnamefont {A.}~\bibnamefont {Surzhykov}},\ }\href {\doibase
  10.1103/PhysRevA.102.012812} {\bibfield  {journal} {\bibinfo  {journal}
  {Phys. Rev. A}\ }\textbf {\bibinfo {volume} {102}},\ \bibinfo {pages}
  {012812} (\bibinfo {year} {2020})}\BibitemShut {NoStop}%
\bibitem [{\citenamefont {Schmiegelow}\ \emph {et~al.}(2016)\citenamefont
  {Schmiegelow}, \citenamefont {Schulz}, \citenamefont {Kaufmann},
  \citenamefont {Ruster}, \citenamefont {Poschinger},\ and\ \citenamefont
  {Schmidt-Kaler}}]{Schmiegelow2016ionabsorbption}%
  \BibitemOpen
  \bibfield  {author} {\bibinfo {author} {\bibfnamefont {C.}~\bibnamefont
  {Schmiegelow}}, \bibinfo {author} {\bibfnamefont {J.}~\bibnamefont {Schulz}},
  \bibinfo {author} {\bibfnamefont {H.}~\bibnamefont {Kaufmann}}, \bibinfo
  {author} {\bibfnamefont {T.}~\bibnamefont {Ruster}}, \bibinfo {author}
  {\bibfnamefont {U.}~\bibnamefont {Poschinger}}, \ and\ \bibinfo {author}
  {\bibfnamefont {F.}~\bibnamefont {Schmidt-Kaler}},\ }\href {\doibase
  10.1038/ncomms12998} {\bibfield  {journal} {\bibinfo  {journal} {Nature
  Communications}\ }\textbf {\bibinfo {volume} {7}},\ \bibinfo {pages} {12998}
  (\bibinfo {year} {2016})}\BibitemShut {NoStop}%
\bibitem [{\citenamefont {C.~Cohen-Tannoudji}\ and\ \citenamefont
  {Grynberg}(1998)}]{Tannoudji}%
  \BibitemOpen
  \bibfield  {author} {\bibinfo {author} {\bibfnamefont {J.~D.-R.}\
  \bibnamefont {C.~Cohen-Tannoudji}}\ and\ \bibinfo {author} {\bibfnamefont
  {G.}~\bibnamefont {Grynberg}},\ }\href@noop {} {\emph {\bibinfo {title}
  {Atom-Photon Interactions: Basic Processes and Applications}}}\ (\bibinfo
  {publisher} {John Wiley \& Sons},\ \bibinfo {year} {1998})\BibitemShut
  {NoStop}%
\end{thebibliography}%



\end{document}